\documentclass[paper]{JHEP3}
\usepackage{epsfig}
\def    \be             {\begin{equation}}
\def    \ee             {\end{equation}}
\def    \ba             {\begin{eqnarray}}
\def    \ea             {\end{eqnarray}}

\def    \=              {\;=\;}
\def    \frac           #1#2{{#1 \over #2}}

\def    \bra#1          {\mbox{$\langle #1 |$}}
\def    \ket#1          {\mbox{$| #1 \rangle$}}

\def    \gev            {\mbox{$\mathrm{GeV}$}}
\def\mub {\mbox{$\mu\mathrm{b}$}}

\def    \pt             {\mbox{$p_T$}}

\newcommand     \MSB            {\ifmmode {\overline{\rm MS}} \else 
                                 $\overline{\rm MS}$  \fi}
\def    \muf            {\mbox{$\mu_{\rm F}$}}

\def    \mur            {{\mbox{$\mu_{\rm R}$}}}

\def    \as             {\ifmmode\alpha_s\else$\alpha_s$\fi}

\def \oacube {\mbox{$ {\cal O}(\alpha_s^3)$}}

\def\rt1{\raisebox{-1ex}{\rlap{$\; \rho \to 1 \;\;$}}
\raisebox{.4ex}{$\;\; \;\;\simeq \;\;\;\;$}}

\def \sss{\scriptscriptstyle}
\def \ss{\scriptstyle}
\def \lsim{\mathrel{\vcenter
     {\hbox{$<$}\nointerlineskip\hbox{$\sim$}}}}
\def \gsim{\mathrel{\vcenter
     {\hbox{$>$}\nointerlineskip\hbox{$\sim$}}}}

\def\ppbar{\mbox{$p \bar{p}$}}
\def\met{$\rlap{\kern.2em/}E_T$}

\title{QCD analysis of first $b$ cross section data at 1.96~TeV}
\vfill                                                       
\author{Matteo Cacciari \\
Dipartimento di Fisica, Universit\`a di Parma, Italy, and\\
LPTHE, Universit\'e P. et M. Curie (Paris 6), France~\footnote{Present address}\\
  E-mail: \email{cacciari@lpthe.jussieu.fr}}
\author{Stefano Frixione \\ 
INFN, Sezione di Genova, Italy\\
  E-mail: \email{Stefano.Frixione@cern.ch}}
\author{Michelangelo L. Mangano \\
CERN, Theoretical Physics Division, Geneva, Switzerland\\
  E-mail: \email{Michelangelo.Mangano@cern.ch}}
\author{Paolo Nason \\
INFN, Sezione di Milano, Italy\\
  E-mail: \email{Paolo.Nason@mib.infn.it}}
\author{Giovanni Ridolfi \\
INFN, Sezione di Genova, Italy\\
  E-mail: \email{Giovanni.Ridolfi@ge.infn.it}}
\abstract{ The first data on bottom quark production in $p\bar{p}$
collisions at 1.96~TeV have recently been obtained by the CDF
collaboration. These data probe the region of $ \pt\sim 0$, providing
a new invaluable input on the issue of the compatibility between
next-to-leading-order (NLO) QCD and data. We reconsider the evaluation
of the $b$ cross section, in view of recent theoretical developments,
and of the latest inputs on structure function fits.  We show that the
new CDF measurements are in good agreement with NLO QCD.  If CDF
preliminary data are confirmed, a long-standing discrepancy between
NLO QCD predictions and hadron-collider data can be settled.}
\preprint{CERN-TH/2003-298 \\ Bicocca-FT-03-33 \\GEF-TH-14/2003
	 \\LPTHE-03-37 \\UPRF-2003-31 \\hep-ph/0312132}

\begin{document}

\section{Introduction}

The measurement of the bottom quark production cross section in
\ppbar\ collisions has provided for the past fifteen years one of the most
significant challenges to the ability of perturbative QCD to accurately
predict absolute rates in hadronic collisions.  Measurements of the
transverse momentum (\pt) spectrum in the region $\pt > m_b$ (the bottom quark
mass)  have been
performed by the UA1 experiment~\cite{Albajar:1987iu} at the
S$\bar{p}p$S ($\sqrt{S}=630$~GeV) and by the
CDF~\cite{Abe:1993sj,Abe:1995dv,Acosta:2001rz} 
and D0~\cite{Abachi:1994kj} experiments at the
Tevatron ($\sqrt{S}=1.8$~TeV). Comparisons of Tevatron data with 
next-to-leading-order (NLO, i.e. $\oacube$)
predictions~\cite{Nason:1988xz,Nason:1989zy} have shown a systematic
excess. The precise size of this excess depends on the input
parameters used for the theoretical calculation, which is affected by
an uncertainty of up to 50\% due to the choice of renormalization and
factorization scales (\mur, \muf) and by additional uncertainties due
to the choice of parton distribution functions (PDFs) and of the value of
the $b$ quark mass.  Nevertheless, the central value of the NLO
prediction has typically been quoted as being smaller than the data by
factors varying between 2 and 3.

Aside from the radical and interesting hypothesis that physics beyond
the standard model is at work~\cite{Berger:2000mp}, the source of this
discrepancy in the context of QCD has been searched for in various
directions. At the perturbative level, the large scale dependence at
NLO is a symptom of large higher-order contributions. First of all it
is well known that there are new partonic processes which appear first
at \oacube (such as gluon splitting). Furthermore, there are large
logarithmic corrections which are present at all orders of
perturbation theory.  These can arise from several sources: on
one side there are logarithms of the ratio of the hadronic center of
mass energy and the quark
mass~\cite{Nason:1988xz,Collins:1991ty,Catani:1990eg} (the so-called
small-$x$ effects, $x\sim m_b/\sqrt{S}$). On the other, multiple gluon
radiation leads at large \pt\ to towers of logarithms of
$\pt/m_b$~\cite{Cacciari:1993mq}.  At the non-perturbative level, it
has been noted~\cite{Mangano:1997ri} that the \pt\ spectrum of $b$
hadrons ($H_b$) in hadronic collisions has a large sensitivity to the
parameterization of the $b\to H_b$ fragmentation function,
$D_b(z)$. Even assuming the applicability of the factorization
theorem, an assumption which at low \pt\ remains to be validated, it
is therefore crucial to ensure that the extraction of $D_b(z)$ from
$e^+e^-$ data is performed in a fashion consistent with its
application in the context of hadronic collisions, an issue often
overlooked in the past.

The quantitative analysis of small-$x$ effects in $b$ production has
followed several types of approaches.  Some of
them~\cite{Levin:1991ry} do not attempt to include the exact NLO
results, thus leading to a radical departure from the QCD-improved
parton model. These approaches generally lead to very large $K$
factors for heavy flavour production, even at high \pt, where these
effects should be reduced. Large effects, strongly dependent on the
chosen fit of unintegrated gluon densities, are also found in the
small-$x$ MC implementation of Jung~\cite{Jung:2001rp}.  The approach
of Collins and Ellis~\cite{Collins:1991ty} aims instead at computing
the small-$x$ enhanced effects that are \emph{not already included} in
the NLO results. In this approach one finds at the Tevatron 
corrections not larger than
20-30{\%}.

The resummation of the logarithms of $\pt/m_b$, with next-to-leading
logarithmic accuracy (NLL), and the matching with the fixed-order,
exact NLO calculation for massive quarks, has been performed in
\cite{Cacciari:1998it} (FONLL).\footnote{The matching with the massive
result at low \pt\ is essential, due to the large size of mass
corrections up to $\pt\sim 20$~GeV. 
Lack of mass effects~\cite{Binnewies:1998vm} will therefore
erroneously overestimate the production rate at small \pt.} A
calculation with this level of accuracy is also available for $b$
production in $e^+e^-$ collisions~\cite{Nason:1999zj},
 and has been used for the
extraction of non-perturbative fragmentation
functions from LEP and SLC
data~\cite{Heister:2001jg}.  The equivalence of the perturbative
inputs allows one to consistently convolute these functions with the
FONLL $b$-quark spectra in hadronic collisions, leading to FONLL
predictions for the $H_b$ spectrum.  A comparison of these
predictions with CDF data at 1.8~TeV for $B^\pm$-meson production in the
range $6~\gev<\pt<20$~GeV has been presented in~\cite{Cacciari:2002pa}.
There the ratio between data and theory,  averaged over the given
\pt\ range, was reduced to a factor of
1.7, compatible with the residual theoretical
and experimental uncertainties.
This finding is consistent
with experimental evidence from D0 (see the last paper
in~\cite{Abachi:1994kj}) that the inclusive rate of
jets containing $b$ quarks -- a quantity largely insensitive to the details of
the perturbative and non-perturbative fragmentation -- agrees with NLO
QCD~\cite{Frixione:1996nh}. 

  Non-perturbative fragmentation effects are expected to play a much
reduced role in the prediction of the total production rate, 
as they only smear the $p_T$ spectrum of the quark.
Furthermore, 
effects due to  initial-state multiple-gluon emission average to 0 after
\pt\ integration.
It follows that the measurement of $b$ production down to $\pt \sim 0$
provides a crucial input for clarifying the origin of the residual
discrepancy between QCD and data: an improved agreement would
strongly support the idea that the blame rests on our incomplete
understanding of the fragmentation phase. A residual, or increased,
disagreement, would support the relevance of small-$x$ effects, which
are expected to grow at low \pt.  For this reason, the recent CDF
release~\cite{cdfrun2} 
of preliminary data on  $H_b$ production at $\sqrt{S}=1.96$~TeV
in the domain $\pt>0$, $\vert y_{H_b} \vert < 0.6$ provides us with a new,
crucial input.  In this paper we therefore re-evaluate the
theoretical predictions developed in~\cite{Cacciari:2002pa}, 
extending the range in
\pt\ down to 0, and reviewing the 
theoretical systematics.
In addition, we compare these results with those
obtained using the MC@NLO code~\cite{Frixione:2003ei}, which
merges the full NLO matrix elements with the complete shower evolution
and hadronization performed by the {\small HERWIG} Monte Carlo. As discussed in
detail in~\cite{Frixione:2003ei}, this comparison probes a few features
where FONLL and MC@NLO differ by 
effects beyond NLO: the evaluation
of subleading logarithms in higher-order emissions, in particular in
the case of gluon emission from the $b$ quark, and the hadronization
of the heavy quark, which in MC@NLO is performed through {\small
HERWIG}'s cluster model, tuned on $Z^0\to H_b X$ decays.

\section{Theoretical results and uncertainties}

We start by considering the integrated cross section for $b$ quarks,
reported by CDF in the domain $\pt>0$, $\vert y_b \vert < 1 $. This
depends only indirectly on the
fragmentation and on the resummation of \pt\ logarithms, so we start by
quoting the results obtained with the standard NLO calculation. The
total cross section is defined by the integral of the single-inclusive
\pt\ distribution, with renormalization and factorization scales
defined by $\mu_{R,F}=\xi_{R,F}\mu_0$, where $\mu_0^2=\pt^2+m_b^2$.
The central values of our predictions are obtained with $\xi_{R,F}=1$
and $m_b=4.75$~GeV.  
To avoid the accidental compensation between the \muf\ and the \mur\
dependence of the cross section occurring when the two scales are
kept equal, we compute the scale uncertainty by varying
\mur\ and \muf\ independently over the range $0.5<\xi_{R,F}<2$,
with the constraint $0.5 <\xi_R/\xi_F < 2$. 
The mass uncertainty corresponds to the range $4.5~\gev<m_b<5$~GeV.  The
result, using CTEQ6M~\cite{Pumplin:2002vw} as PDF, is:
\be \label{eq:nlocteq}
\sigma^{\rm NLO}_b(\vert y_b \vert < 1) = 
       23.6 \; {{\ss  +4.5\atop \ss -3.6}}_{\sss m_b}
            \; {{\ss  +10.8\atop \ss -6.3}}_{\sss \mu_R,\mu_F}
                                     \; \mu{\rm b} \; .
\ee
For comparison, the full FONLL calculation and MC@NLO lead to slightly
larger central values, 25.0 $\mu$b and 25.2 $\mu$b respectively. This
difference is due to higher-order effects included in these
calculations, which render the rapidity distribution narrower than
 that computed at the NLO.

\begin{figure}[t]
\begin{center}
\epsfig{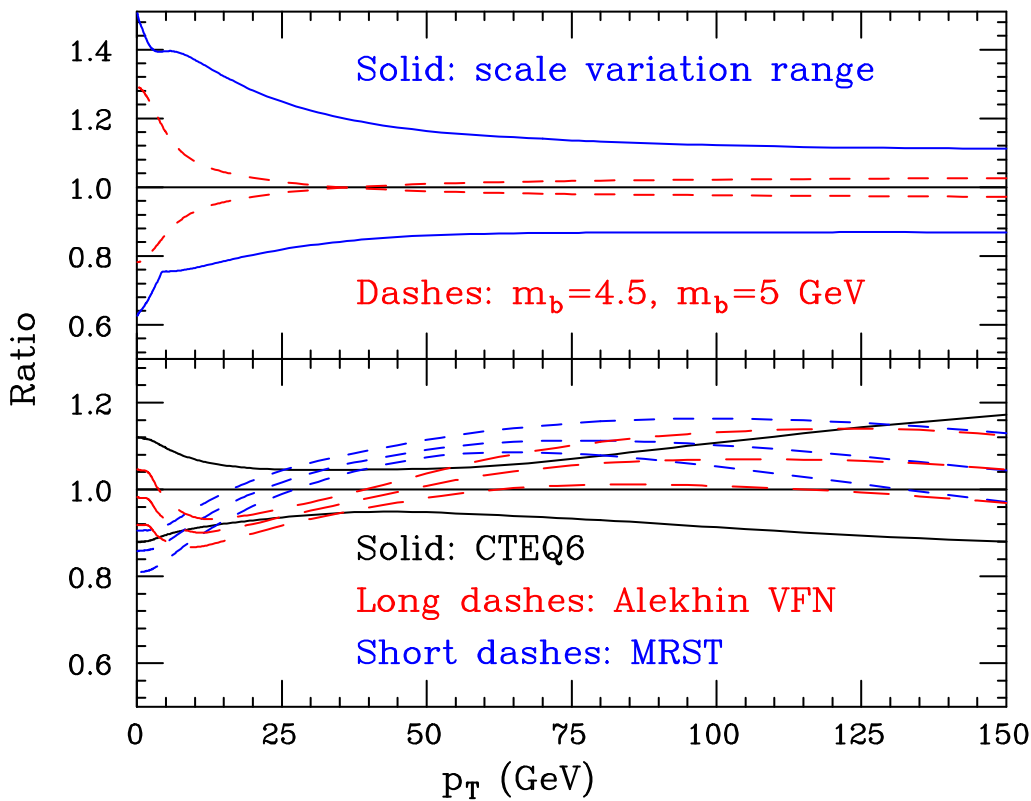}
\end{center}
\caption{Lower panel: variations of the FONLL $b$-quark \pt\ spectrum 
($y_b=0$) due to PDF uncertainties, normalized to the central CTEQ
 prediction. Upper panel:
variation of the spectrum due to scale and mass
  variations, normalized to the default prediction.}
\label{fig:pdfscale}
\end{figure}
We studied the PDF uncertainties using three different sets of fits with
systematics: CTEQ6~\cite{Pumplin:2002vw},
MRST~\cite{Martin:2002aw} and Alekhin~\cite{Alekhin:2002fv}. 
Following the prescriptions in those papers, we get:\footnote{We 
verified that the prescription proposed for CTEQ's sets
in~\cite{Sullivan:2002jt} leads to minimal differences in the results.}
\ba
\sigma^{\rm NLO}_b(\vert y_b \vert < 1, \mbox{CTEQ}) &=&   23.6 \pm
2.3_{\sss {\rm PDF}} \; \mu{\rm b} 
\; , \\ 
\sigma^{\rm NLO}_b(\vert y_b \vert < 1, \mbox{MRST}) &=&   20.8 \pm
0.9_{\sss {\rm PDF}}   \; \mu{\rm b}
\; , \\
\sigma^{\rm NLO}_b(\vert y_b \vert < 1, \mbox{Alekhin, FFN}) &=&
24.3 \pm 1.3_{\sss {\rm PDF}}   \; \mu{\rm b}
\; , \\
\sigma^{\rm NLO}_b(\vert y_b \vert < 1, \mbox{Alekhin, VFN}) &=&
22.4 \pm 1.2_{\sss {\rm PDF}}   \; \mu{\rm b} \; .
\ea
The two Alekhin values refer to the choice of the fixed (FFN) and
variable (VFN)
flavour number schemes. We note that the CTEQ
prediction has an uncertainty twice as large as the others. This is
a known fact, related to the different prescriptions used to define the 
error ranges. A similar result holds in fact 
for other observables~\cite{Stump:2003yu}.
Secondly, we note that the MRST central value is only barely 
consistent with the others, even within the error bars. 

\begin{figure}[t]
\begin{center}
\epsfig{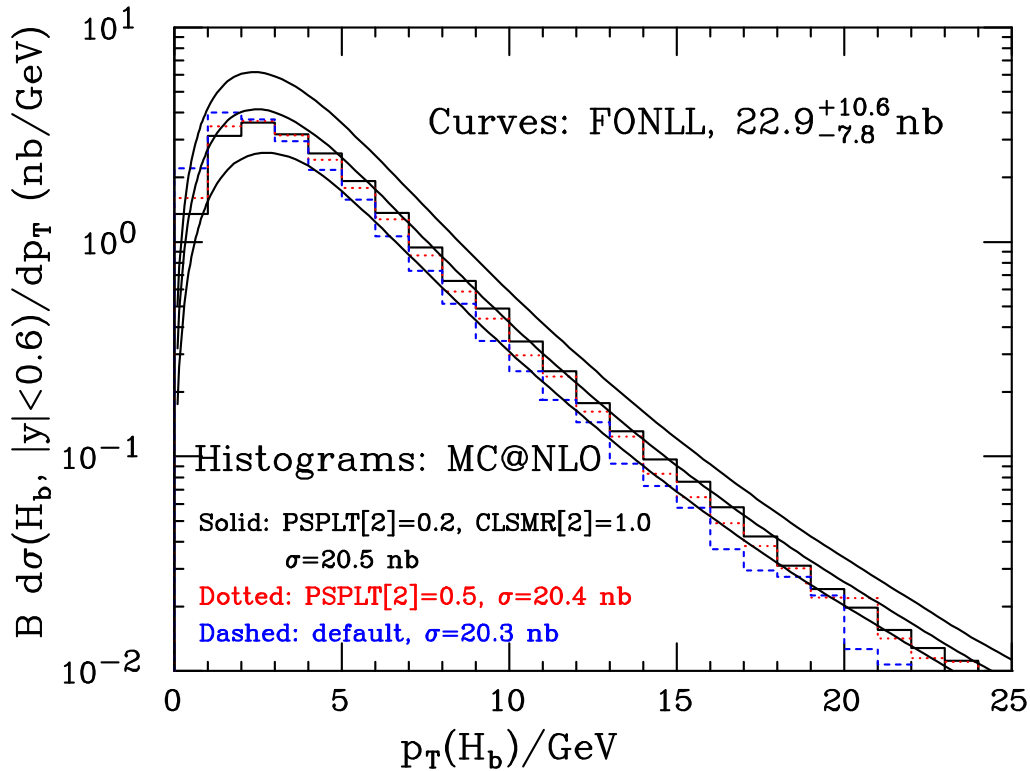}
\end{center}
\caption{The $H_b$ spectrum for $\vert y_{H_b} \vert < 0.6$. $H_b$ 
stands for a $b$ or $\bar{b}$
flavoured hadron. The cross section has been multiplied
by $B \equiv BR(H_b \to J/\Psi \to \mu^+\mu^-) = 6.82\times10^{-4}$ 
for future convenience.
The solid curves
  give the FONLL overall uncertainty band as described in the text. The
  central solid curve corresponds to the central value, namely that 
  obtained with CTEQ6M, $\mu=\mu_0$, and $m_b=4.75$~GeV. The histograms
  present MC@NLO results obtained with different $b$ hadronization parameters.}
\label{fig:Hbpt}
\end{figure}
We move now to the FONLL analysis of the \pt\ spectra.
The lower panel of Fig.~\ref{fig:pdfscale} 
shows the PDF uncertainty of the $b$-quark \pt\
spectrum at $y_b=0$. We plot the upper and lower results obtained
using the CTEQ6, MRST and Alekhin fits, normalized to the central
CTEQ6M prediction. In all cases, $m_b=4.75$ GeV and
$\xi_{R,F}=1$. The spectrum confirms a poor consistency between  
MRST results and the others. 
Only for $\pt>100$~GeV the three bands are consistent with each other.
In the region below $\sim 50$~\gev, the theoretical uncertainty is
however dominated by the effects of scale and mass variation, which
are shown in the upper panel of Fig.~\ref{fig:pdfscale}. 
The scale variation is obtained by varying \mur\ and \muf\ as was done for
the total rates. 
What we show is the envelope of the upper 
and lower results. The points on the 
curves therefore do not necessarily all correspond to the same  scale
choice, and the shape is not representative of a specific scale choice.

We summarize these results in Fig.~\ref{fig:Hbpt}, while at the same
time going from the quark to the hadron level. The solid curves
give the error band for the $\pt(H_b)$ spectrum as predicted by FONLL by 
summing in quadrature the scale, mass and PDF uncertainties. We selected 
CTEQ6M for our central prediction, and applied a $10\% $ PDF uncertainty in
each \pt\ bin, to reflect the effects shown in
Fig.~\ref{fig:pdfscale}. The inclusion of non-perturbative effects
related to the $b\to H_b$ fragmentation has been performed according to
the framework described in~\cite{Cacciari:2002pa}.
The momentum fraction $z$ (see below) used in this approach for scaling the $b$
quark momentum to the $H_b$ hadron one is extracted 
from a normalized Kartvelishvili et al. distribution, 
$D(z) = (\alpha+1)(\alpha+2) x^\alpha (1-x)$~\cite{Kartvelishvili:1977pi}. 
The phenomenological parameter $\alpha$ is fixed to 25.6, 29.1, 34 for  $m_b$ =
4.5, 4.75 and 5 GeV respectively by tuning it to moments of $e^+e^-$ data (see
\cite{Cacciari:2002pa}).  We emphasize that these values for $\alpha$ depend 
also on other details
and parameters of the perturbative calculation (like, for instance, the value
of the strong coupling $\alpha_s$), and cannot therefore simply be used with
other $b$-quark perturbative spectra without retuning them to $e^+e^-$ data.

In Fig.~\ref{fig:Hbpt} we also present the
results obtained with MC@NLO, using default values of mass, scales and
PDF,\footnote{We verified that mass, scale, and PDF uncertainties in 
MC@NLO are of the same size as those in FONLL.} and for various choices 
of the $b$ hadronization parameters, as described in~\cite{Frixione:2003ei}. 
The dashed histogram corresponds to the {\small HERWIG} default, whereas
the solid (dotted) histogram has been obtained by setting 
{\small PSPLT(2)}=0.2 and {\small CLSMR(2)}=1 ({\small PSPLT(2)}=0.5).
The agreement with the central FONLL prediction is satisfactory in
terms of shape and rate in the case of the solid histogram; the other
two choices of $b$ hadronization parameters are seen to return softer
$\pt(H_b)$ spectra.

\begin{figure}[t]
\begin{center}
\epsfig{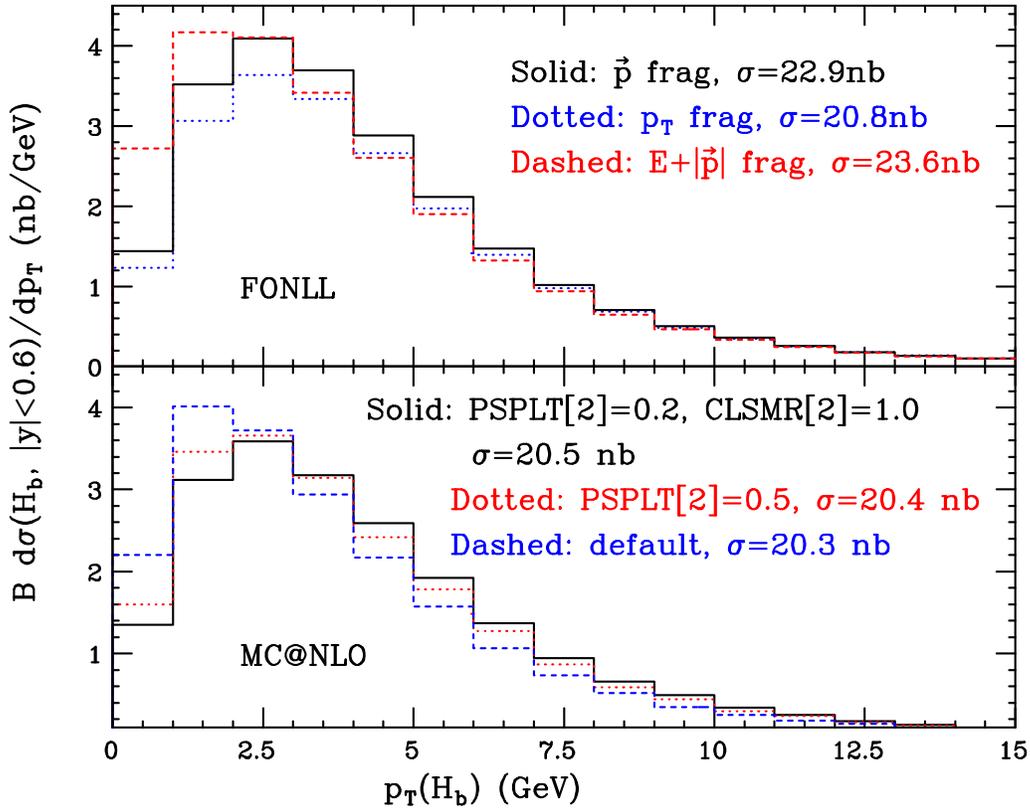}
\end{center}
\caption{Dependence of the $\pt(H_b)$ spectrum on the choice of
  fragmentation prescription in FONLL (upper panel), and on the $b$ 
  hadronization parameters in MC@NLO (lower panel, same curves as in
  Fig.~\protect\ref{fig:Hbpt}).}
\label{fig:bfragm}
\end{figure}
Different choices of the $b$ hadronization parameters in {\small HERWIG}
correspond to different treatments of fragmentation in FONLL; the latter
give additional uncertainties with respect to those, of purely perturbative
origin, that we discussed so far. Here, we shall not study the fragmentation
uncertainties systematically, and we shall limit ourselves to illustrate the 
typical sizes of these effects. As a result, we shall not cumulate the effect 
of the following variations with those above.
The choice of the fragmentation variable for the $b\to H_b$
transition, which at high \pt\ is
irrelevant, may lead to some differences at small \pt, where
non-factorizable effects can be significant. 
The upper panel of Fig.~\ref{fig:bfragm} shows the effect of three separate
choices of fragmentation. The solid line is the 
default used in this work, $\vec p(H_b) =z \vec p(b)$, the three-momenta
being taken in the laboratory frame (the curves in 
Fig.~\ref{fig:Hbpt} have been obtained with this choice).
The dotted line corresponds to 
$\pt(H_b)=z\pt(b)$ and $y_{H_b}=y_b$.  The dashed line to
$(E+p)_{H_b}=z(E+p)_{b}$.  
For comparison, the lower panel of Fig.~\ref{fig:bfragm} presents the 
MC@NLO results already shown in Fig.~\ref{fig:Hbpt}.
Different fragmentation/hadronization choices lead to 
different shifts in $y_{H_b}$, which can have an impact on the cross 
section within the $\vert y_{H_b} \vert < 0.6$ range when $\pt\lsim m_b$. 
For FONLL, these shifts modify the total rate at the level of $\sim 10$\%, 
while no significant change is observed with the MC@NLO hadronization.
This is not surprising, since the fragmentation mechanism is strictly
speaking only applicable in the large-$\pt$ region, while the hadronization
embedded into parton shower Monte Carlos should work for any $\pt$'s (and
thus requires more tuning to data).

\section{Comparison with experimental data}

\begin{figure}[t]
\begin{center}
\epsfig{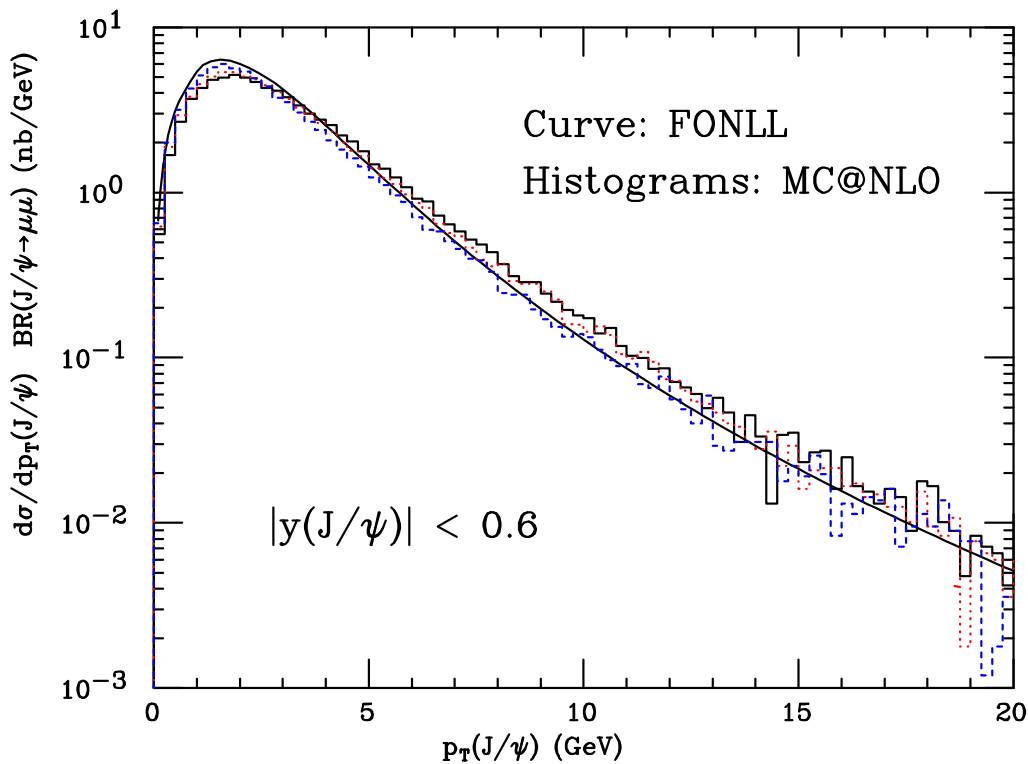}
\end{center}
\caption{FONLL prediction for the $J/\psi$ spectrum (solid curve), for 
central values of mass, scales, PDF, and fragmentation scheme, compared
to MC@NLO results (histograms).}
\label{fig:jpsi-mqmb}
\end{figure}

We present now the comparison of our results with data.  Since the CDF
measurement is performed using inclusive $B\to J/\psi+X$ decays, with
$J/\psi\to\mu^+\mu^-$, we first quote results for this observable. 
The FONLL prediction uses the $B\to J/\psi+X$
spectra measured by BaBar and CLEO~\cite{Aubert:2002hc}, applied
to all $H_b$ states.  
We verified that the
$\pt(J/\psi)$ distribution shows little sensitivity to the input choice for
the decay spectrum. On the other hand, MC@NLO generates the 
$B\to J/\psi+X$ decays using the perturbative $b\to c $ decay
spectrum. The resulting $\pt(J/\psi)$ spectra are compared in
Fig.~\ref{fig:jpsi-mqmb}. At variance with the case of Fig.~\ref{fig:Hbpt},
the $J/\psi$ spectra predicted by MC@NLO are generally harder than that
predicted by FONLL. The best agreement is obtained with the choice of the
$b$ hadronization parameters that gives the worst agreement in the case
of the $\pt(H_b)$ spectrum. Since the $B\to J/\psi$ fragmentation in
FONLL is obtained directly from data, this points out that a more 
realistic treatment of the $B$ decay would be needed in {\small HERWIG}.
Clearly, this effect can be eliminated by replacing the internal 
{\small HERWIG} $B$-decay routines with standard $B$-decay packages.

Defining $\sigma_{J/\psi}= \sigma(H_b\to J/\psi,
\pt(J/\psi)>1.25, \vert y_{J/\psi} \vert<0.6) 
\times {\rm BR}(J/\psi\to\mu^+\mu^-) $, 
where~\cite{Hagiwara:fs}
\begin{eqnarray*}
{\rm BR}(J/\psi\to\mu^+\mu^-) = 5.88 \times 10^{-2}
\end{eqnarray*}
and, to be used below, 
\begin{eqnarray*}
&&{\rm BR}(H_b\to J/\psi) = 1.16\times 10^{-2}\; ,\\
&&B \equiv {\rm BR}(H_b\to J/\psi\to\mu^+\mu^-) = 6.82\times 10^{-4}\; ,
\end{eqnarray*} 
CDF's result is:
\be
\sigma_{J/\psi}^{\rm CDF}= 19.9 \; {\ss +3.8 \atop \ss -3.2}_{stat+syst} \; 
\mbox{nb}\;,
\ee
in excellent agreement with ours:
\be
\sigma_{J/\psi}^{\rm FONLL} = 18.3 \; {\ss +8.1 \atop \ss -5.7}\; \mbox{nb}
\; ,
\ee
where we combined in quadrature the uncertainties from scale, mass and
PDF variations.
CDF then deconvolutes the $H_b\to J/\psi$ decay, quoting the following
result for
$\sigma_{H_b}=
\sigma(H_b, \vert y_{H_b} \vert<0.6) \times B$:
\be
\sigma_{H_b}^{\rm CDF}   =   24.5 \; 
       {\ss +4.7 \atop \ss -3.9}_{stat+syst} \; \mbox{nb}\;,
\ee
in good agreement with  our estimate:
\be
\sigma_{H_b}^{\rm FONLL} = 22.9 \;  {\ss +10.6 \atop \ss -7.8} \mbox{nb}
\; .
\ee
After dividing the $H_b$ rate by a factor $B\times 0.61$ 
to correct for the branching ratio and for the different $y_b$ 
range,\footnote{The correction factor for the $y_b$ range predicted
by FONLL and MC@NLO is also 0.61.} CDF gives a $b$-quark
cross section for $\vert y_b \vert < 1$:
\be
\sigma_b^{\rm CDF}(\vert y_b \vert < 1) 
   =   29.4 \; {\ss +6.2 \atop \ss -5.4}_{stat+syst} \; \mu\mbox{b}\;,
\ee
to be compared with our estimate (Eq.~(\ref{eq:nlocteq}) with the
additional 10\% PDF uncertainty, rescaled to the FONLL result):
\be
\sigma_b^{\rm FONLL}(\vert y_b \vert < 1) 
   =   25.0
       \; {\ss +12.6 \atop \ss -8.1} \; \mu\mbox{b}\;.
\ee
We note that the difference between the central values of data and
theory increases from 5\% to 15\% when going from the result closest to
the direct experimental measurement (the $J/\psi$ cross section) to
those which require more deconvolution and acceptance corrections
(the inclusive $b$ quark rate). This seems to indicate intrinsic
differences, to be ultimately considered as part of the systematic
error, in the modeling of the transition from the quark to the hadrons,
and vice versa. Effects of this size are consistent with what
we showed in Fig.~\ref{fig:bfragm}.

\begin{figure}[t]
\begin{center}
\epsfig{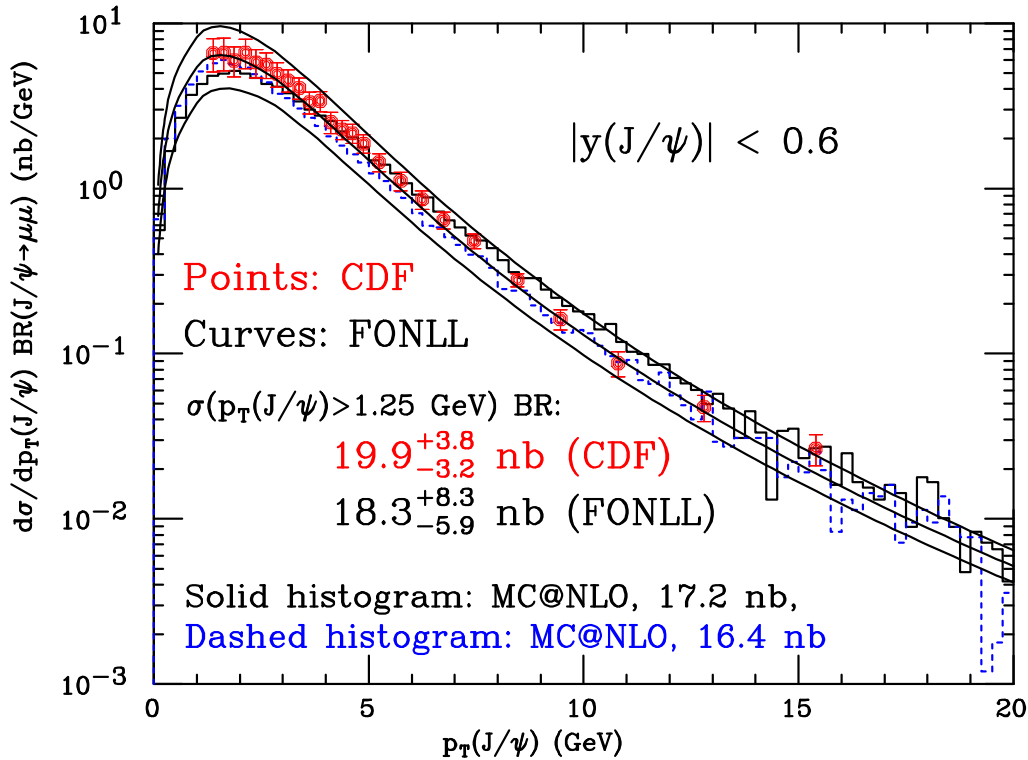}
\end{center}
\caption{CDF $J/\psi$ spectrum from $B$ decays. 
The theory band represents the FONLL systematic uncertainties,
propagated from Fig.~\ref{fig:Hbpt}. Two MC@NLO predictions are
also shown (histograms), with the same patterns as in Fig.~\ref{fig:bfragm}.}
\label{fig:cdfpsi}
\end{figure}

We finally present in Fig.~\ref{fig:cdfpsi} our prediction for the
$J/\psi$ spectrum, obtained by convoluting the FONLL result with the
$J/\psi$ momentum distribution in inclusive $B\to J/\psi+X$
decays.\footnote{An earlier version of this work had a curve with a
  slightly different slope. We correct here an accidental error in the
  treatment of the $H_b$ decay: in the previous version the $b$-quark
  mass, instead of 
  the $b$-hadron mass (which we take equal to 5.3 GeV), 
  was used in the boost to the $H_b$ rest frame before its decay.}
The data lie well within the uncertainty band, and are in very good
agreement with the central FONLL prediction. We also show the two
MC@NLO predictions corresponding to the two extreme choices of the
$b$ hadronization parameters considered in this work; very good 
agreement with data is obtained for one of them in terms of shape, 
with the normalization being slightly low (still within $1\sigma$ 
of the mass and scale uncertainties).

\begin{figure}[t]
\begin{center}
\epsfig{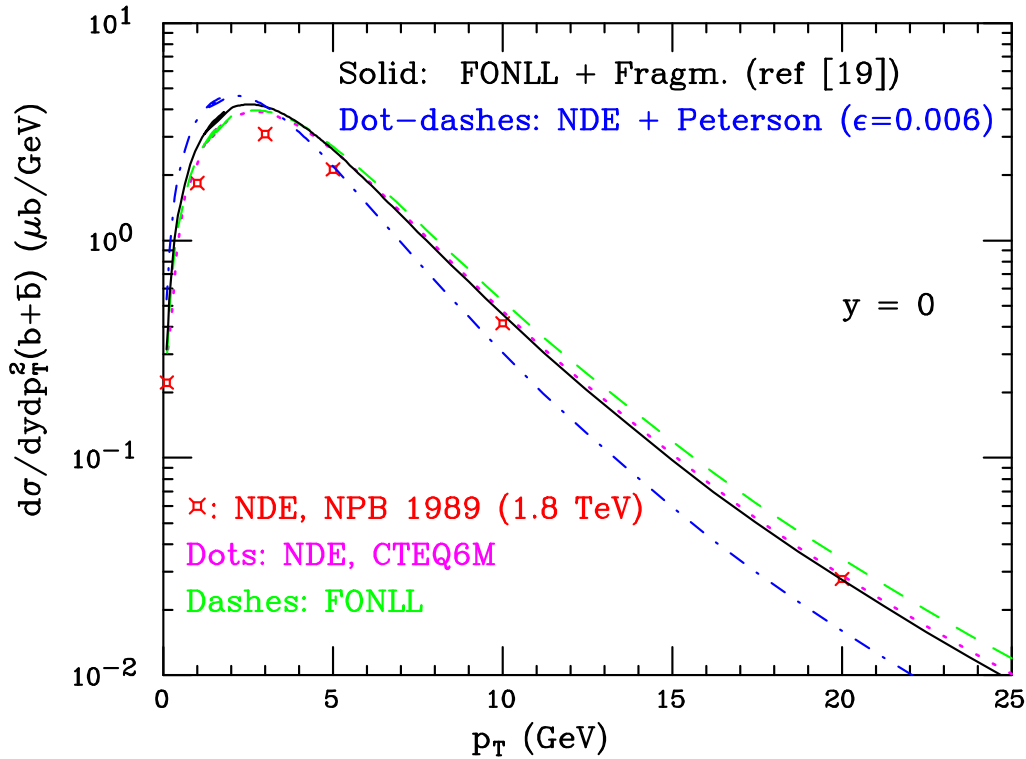}
\end{center}
\caption{Evolution of the NLO QCD predictions over time, for $\sqrt{S} =
1800$ GeV. }
\label{fig:history}
\end{figure}

We stress that both FONLL and MC@NLO are based on the NLO result
of~\cite{Nason:1989zy} (henceforth referred to as NDE), 
and only marginally enhance the cross section
predicted there, via some higher-order effects.  The most relevant
change in FONLL with respect to old predictions lies at the
non-perturbative level, i.e. in the treatment
of the $b\to H_b$ hadronization, which makes use~\cite{Cacciari:2002pa}
of the
moment-space analysis of the most up-to-date data on $b$ fragmentation
in $e^+e^-$ annihilation.  The evolution of the NLO theoretical
predictions over time is shown in Fig.~\ref{fig:history}. Here
we plot the original central prediction of NDE for $\sqrt{S}=$1.8~TeV 
(symbols), obtained using NLO QCD partonic cross sections convoluted with 
the PDF set available at the time, namely DFLM260~\cite{Diemoz:1987xu}. 
The same calculation, performed with the CTEQ6M
PDF set (dotted curve),
shows an increase of roughly 20\% in rate in the region
$\pt<10$~GeV. The effect of the inclusion of the resummation of NLL
logarithms is displayed by the dashed curve, and is seen to be modest
in the range of interest. Finally, we compare the
original NDE prediction after convolution with the Peterson
fragmentation function ($\epsilon=0.006$, dot-dashed curve), with the
FONLL curve convoluted with the fragmentation function extracted
in~\cite{Cacciari:2002pa} (solid curve). 
Notice that the effect of the fragmentation obtained in~\cite{Cacciari:2002pa} 
brings about a modest decrease of the cross section (the difference 
between the dashed and solid curves), 
while the traditional Peterson fragmentation with 
$\epsilon=0.006$ has a rather pronounced effect (the difference 
between the symbols and the dot-dashed curve).
Thus, the dominant change in the theoretical prediction 
for heavy flavour production from the original NDE calculation 
up to now appears to be the consequence of more precise 
experimental inputs to the bottom fragmentation
function~\cite{Heister:2001jg}, that have shown that non-perturbative 
fragmentation effects in bottom production are much smaller 
than previously thought.

\begin{figure}[t]
\begin{center}
\epsfig{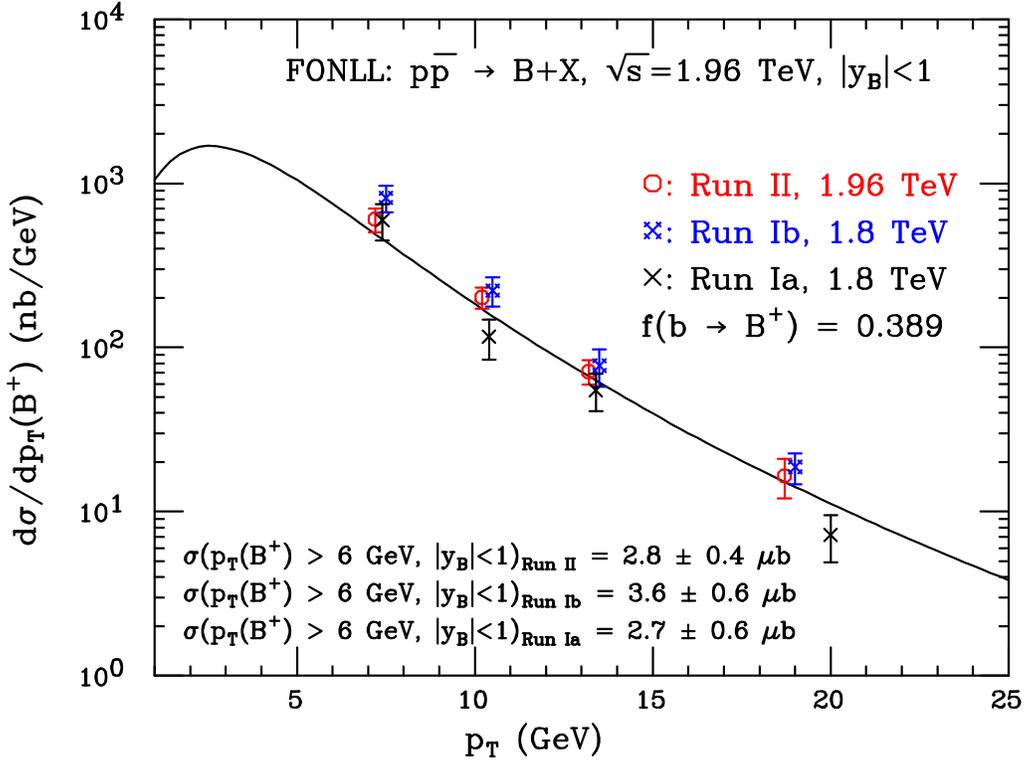}
\end{center}
\caption{Evolution of the CDF data for exclusive $B^\pm$ production:
  Run~IA\protect\cite{Abe:1995dv}, Run~IB~\protect\cite{Acosta:2001rz} and
  Run~II\protect\cite{cdfrun2}.}
\label{fig:bplus}
\end{figure}

We recall once more that, at the level of central values and neglecting
the theoretical and experimental uncertaintites, 
the FONLL result~\cite{Cacciari:2002pa},
using CTEQ5M PDF, underestimates CDF's run~IB $B^\pm$
rate~\cite{Acosta:2001rz} for $\pt\gsim 6$~GeV by a factor of
1.7. The main improvement in the comparison
between Run~II data and theory comes from the new CDF data, which tend
to be lower than one would have extrapolated from the latest
measurements at 1.8~TeV. To clarify this point, we collect in
Fig.~\ref{fig:bplus} the experimental results from the CDF
measurements of the $B^\pm$ cross section in Run~IA~\cite{Abe:1995dv},
in Run~IB~\cite{Acosta:2001rz} and in Run~II. The rate in the
bin which in the
past showed the worse discrepancy, namely the first bin corresponding to
$\pt(B^\pm)>6$~GeV, evolved from $2.7\pm0.6~\mub$ (Run~IA) to
$3.6\pm0.6~\mub$ (Run~IB), and decreased to  $2.8\pm0.4~\mub$ in
Run~II. The increase in the c.m. energy should have instead led to an
increase by 10-15\%. The Run~II result is therefore lower than the extrapolation
from Run~IB by approximately 30\%. By itself, this result alone would
reduce the factor of 1.7 quoted in~\cite{Cacciari:2002pa} to 1.2 at
$\sqrt{S} = 1.96$~TeV. In addition, the results presented in this
paper lead to an increase in rate relative to the calculation 
of~\cite{Cacciari:2002pa}
by approximately 10-15\%, due to the
change of PDF from CTEQ5M to CTEQ6M.
The combination of these two effects, dominated in
any case by the lower normalization of the new data, leads to the
excellent agreement observed in our work.
We then conclude that the improved agreement between the Run~II
measurements and perturbative QCD is mostly a consequence of improved
experimental inputs (which include up-to-date $\as$ and PDF
determinations).

\section{Conclusions}

In summary, the recent CDF measurement of total $b$-hadron production rates in
$p\bar{p}$ collisions at $\sqrt{S}=1.96$~TeV is in good agreement with
NLO QCD, 
the residual discrepancies being  well
within the  uncertainties due to the choice of scales
and, to a lesser extent, of mass and PDF. A
similar conclusion is reached for the \pt\ spectrum, where the
calculation is improved by the inclusion of the NLL resummation of
collinear logarithms. The improvement in the quality of the
agreement between data and theory relative to previous studies is the
result of several small effects, ranging from a better knowledge of
fragmentation and structure functions and of \as, which constantly
increased in the DIS fits over the years, to the fact that these data
appear to lead to cross sections slightly lower than one would have
extrapolated from the measurements at 1.8~TeV.  The current large
uncertainties in data and theory leave
 room for new physics. However there is no
evidence now that their presence is required for the
description of the data, and furthermore the recent results
of~\cite{Janot:2004cy}  
rule out the existence of a scalar bottom quark in the range preferred
by the mechanism proposed in~\cite{Berger:2000mp}. 
The data disfavour the presence of
small-$x$ effects of the size obtained with the approaches of
refs.~\cite{Levin:1991ry}.  They are instead compatible with
the estimates of~\cite{Collins:1991ty}. 
We thus conclude that
approaches to the small-$x$ problem that can exactly include the
NLO corrections (i.e. that can be used to perform a consistently
matched calculation) should be further pursued, also in light of the
recent progress in small-$x$ resummation~\cite{Ciafaloni:2003rd}.  It
is not unlikely that the consistent inclusion of small-$x$ effects
could further reduce the theoretical errors due to scale variation and
PDF uncertainties.  A final confirmation of the CDF data, as well as
the extension of the measurements to the high-\pt\ region
($\pt>40$~GeV), where the theoretical uncertainty from scale
variations is significantly reduced, would provide additional support to
our conclusions.
While our work has no direct impact on other anomalies reported by CDF
in the internal structure and correlations of
heavy-flavoured jets~\cite{Acosta:2001ct}, we do expect that the
improvements relative to pure parton-level calculations present in
the MC@NLO should provide a firmer benchmark for future studies of 
the global final-state stucture of $b\bar{b}$ events.

\vspace{12pt}
\noindent
{\bf Acknowledgments}: We thank M. Bishai, T. Le Compte
and S. Tkaczyk for discussions about the CDF data, and C. Hearty for
providing us with the 
BaBar data relative to the $H_b\to J/\psi$ spectra.

\end{document}